\documentclass{article}

\usepackage[centertags]{amsmath}
\usepackage{amsfonts}
\usepackage{amssymb}
\usepackage{amsthm}
\usepackage{graphicx}
\usepackage{color}
\usepackage{newlfont}
\usepackage{stmaryrd}
\usepackage{mathrsfs}
\usepackage{euscript}
\usepackage[english]{babel}
\usepackage{multirow}
\usepackage{braket}
\usepackage{bbm}

\usepackage{color}

\numberwithin{equation}{section}

\newcommand{\bbC}{{\mathbb C}}

\newcommand{\bbN}{{\mathbb N}}

\newcommand{\bbR}{{\mathbb R}}

\newcommand{\opunit}{\text{1}\kern-0.22em\text{l}}

\newcommand{\out}{_{\text{OUT}}}

\newcommand{\id}{\textrm{d}}

\newcommand{\refeq}[1]{(\ref{#1})}

\begin{document}

\begin{center}{\bf{\Large Active fluctuation symmetries}}
\end{center}
Christian Maes and Alberto Salazar, Instituut voor Theoretische Fysica\\ KU Leuven\\
12 November 2013

\vspace{1cm}
\begin{center}{\bf{\large Abstract}}
\end{center}

In contrast with the understanding of fluctuation symmetries 
for entropy production, similar ideas applied to the time-symmetric 
fluctuation sector have been less explored. 
Here we give detailed derivations of time-symmetric fluctuation 
symmetries in boundary driven particle systems 
such as the open Kawasaki lattice gas and the zero range model. 
As a measure of time-symmetric dynamical activity  over time $T$ we count the difference 
$(N_{\ell} - N_r)/T$ between the number of particle jumps in or out
at the left edge  and those at the right edge  of the system.
We show that this quantity satisfies a fluctuation symmetry 
from which we derive a new Green-Kubo type 
relation.  It will follow then that the system is more active
at the edge connected to the particle reservoir with the largest chemical potential. 
We also apply these exact relations derived for stochastic particle models to a deterministic case, 
the spinning Lorentz gas, where the symmetry relation for the activity is checked 
numerically. \vspace{1 cm}

\section{Introduction}
Fluctuation relations have emerged from an analysis of entropy production 
in driven dissipative processes. 
These are general and non-perturbative relations, something 
which is not so common in nonequilibrium physics. 
There has therefore been a big interest in such 
fluctuation symmetries in recent decades, as pioneered in 
the papers \cite{em,gc}. 
It was found that such symmetries are an expression of local detailed balance, 
implying that the total path-wise entropy flux is the source term of time-reversal breaking 
in the nonequilibrium action governing the dynamical ensemble; see \cite{maes,LS,crooks,time,poincare}. 
In turn, local detailed balance is implied by and refers to the underlying microscopic time-reversibility that governs the contact 
between the system and each (equilibrium) reservoir in the environment \cite{time,leb,kls,har,der,hal}. 
Also the nonequilibrium free energy relations, called Jarzynski 
relation after \cite{jarz}, are of a very similar nature. \\

The present paper takes
some distance from the original works.   We do not concentrate on the traditional dissipative variables but
we add a novel type of fluctuation symmetries 
belonging to the time-symmetric sector 
of a nonequilibrium system.  That was already initiated in \cite{maarten}, but here we update
our understanding with specific models and new Green-Kubo relations.  
Moreover, since then, a new wave of research interest on a specific time-symmetric quantity, 
the dynamical activity, has emerged \cite{chan,vW,woj,fdr}.  This dynamical activity,
or frenesy as called in the context of 
linear response \cite{fdr}, captures essential nonequilibrium kinetic aspects.   In the present paper differences in dynamical activity between the various contacts of the system with the environment arise from the breaking 
of a spatial symmetry that naturally accompanies
the  nonequilibrium situation.
The main result of this paper 
(in Section \ref{da}) thus gives fluctuation symmetries 
in terms of a difference in dynamical activities.

For the plan of the paper, the next section formalizes the central idea. Besides time-reversal symmetry 
we add a second symmetry which can be spatial or internal and that gives rise 
to additional fluctuation symmetries. The standard 
example of a fluctuation symmetry for the entropy flux 
is then reviewed in Section \ref{se}.  New to this case is the relation with the Kubo formula \cite{kubo}
(and not just the Green-Kubo relations) as it also follows from the fluctuation symmetry. 
We then concentrate on the boundary driven Kawasaki and zero range dynamics 
in Section \ref{da}. We derive in these models the  {\it active} 
fluctuation symmetries for differences in dynamical activity (between the left versus right edge of the system). 
These give fluctuation--activity relations, also including  a
Green-Kubo relation for the mentioned dynamical activity.  Interestingly, we are able to derive there 
that the activity (in terms of the number of particles moving in or out of the system) is highest at 
that side of the system which is in contact with the largest chemical potential. Section \ref{slga} 
applies the latter results to the spinning Lorentz gas, a mechanical model, 
where the notion of dynamical activity gets further realization in a classical physics context.  
The computer simulations we present validate our guesses also in the non-Gaussian fluctuation sector.

\section{General observation}\label{gobs}
We start by presenting the formal content of fluctuation symmetries as in \cite{maes}. 
Let us denote quite generally a fluctuating quantity with the variable $X\in \Omega$, 
where $\Omega$ is the space of 
possible outcomes (e.g. path space). This means that the outcome of $X$ changes and 
is uncertain as, in physical terms, its value depends 
on hidden or more microscopic degrees of freedom. 
In addition we consider the presence of certain involutions $\Theta$ and $\Gamma$ on $\Omega$, i.e., 
transformations which are equal to their inverse and that preserve the elementary 
structure of the space $\Omega$ such as the volume element; these involutions are 
also mutually commuting:
$\Gamma^2=\Theta^2=$ Id, $\Theta\Gamma=\Gamma\Theta$. The fact that $\Theta$ and $\Gamma$ are 
commuting ensures that $\Theta\Gamma$ is also an involution. 

There will always be a reference probability law
$P_o$ for $X$ which is both $\Theta-$ and $\Gamma-$invariant; 
$P_o(\Theta X) = P_o(X) = P_o(\Gamma X)$.
\footnote{For mathematical modeling purposes we consider probability distributions 
for $X$ in $\Omega$. This in turn 
also means that the space $\Omega$ is measurable; in other words, 
it supports some elementary structure 
such as used for integration, so that probabilities 
may admit a density function description.} \\

Our main interest is to formulate a probability law 
$P$ on $X$ for the nonequilibrium process. We assume it
has a density with respect to $P_o$:
\begin{equation}\label{act}
\id P(X) = e^{-A(X)}\,\id P_o(X)
\end{equation}
where  the ``action'' $A$ on the paths or trajectories of the system appears. 
For our purposes, this action will be mostly 
explicitly known by setting a specific context. 
For the meaning and use of the nonequilibrium law \eqref{act}, 
let us pretend for a moment that $X$ 
takes a finite number of values or states so that expectations 
$\langle \cdot \rangle$ under $P$ are simply written as finite sums
\begin{equation}
\langle f(X)\rangle = \sum_{x} f(x) P(x) = \sum_{x} f(x) \,e^{-A(x)}\,P_o(x )
\end{equation}
for an observable $f$ and with $P(x)=\int \id P(X)\delta(X - x )$ the probability that $X=x $.  Note for the notation that
$X$ denotes the random variable while $x$ stands for the different values it may take.\\

Now we present our general observation.  Starting from \eqref{act} let 
us define on $\Omega$,
\begin{eqnarray}\label{defs}
S &:=& A\Theta - A\nonumber\\
\mathcal T &:=& A\Theta + A\\
R &:=& A\Theta\Gamma - A\nonumber
\end{eqnarray}
as functions of $X$. In words, $S$ is the time-antisymmetric part of the action 
while $\mathcal{T} $ is 
the time-symmetric complement. With this the action is 
expressed as 
\begin{equation}
A = \frac 1{2} \big( \mathcal{T} - S\big).
\end{equation}
Note also that $2R = \mathcal T \Gamma - \mathcal T + S + S \Gamma$ so that $R$ is 
the antisymmetric part of $\mathcal T$ under $\Gamma$ when $S$ is antisymmetric under $\Gamma$:
\begin{equation}\label{mir}
S\Gamma=-S \Leftrightarrow R = \frac 1{2}\big({\mathcal T}\Gamma - {\mathcal T}\big)
\end{equation}
Observe that the very definitions \eqref{defs} imply the following identities
\begin{eqnarray}
\langle f(\Theta X)\rangle = \sum_{x } f(x ) e^{-A(\Theta x )}\,P_o(x ) = \langle f(X) e^{-S(X)}\rangle\label{gc}\\
\langle f(\Theta\Gamma X)\rangle = \sum_{x } f(x ) e^{-A(\Theta \Gamma x )}\,P_o(x ) = \langle f(X) e^{-R(X)}\rangle\label{un}
\end{eqnarray}
for all functions $f$ on $\Omega$.
From \eqref{un} we also have for $\Theta-$symmetric observables 
$f=f\Theta$ such as $f={\mathcal T}\Gamma - {\mathcal T}$ that
\begin{equation}
\langle f\Gamma \rangle =  \langle f\, e^{-\frac 1{2}\big(\mathcal T \Gamma - \mathcal T\big) - \frac 1{2}\big(S + S \Gamma\big)}\rangle.
\label{afs}
\end{equation}
We refer to \eqref{afs} as an active fluctuation symmetry for reasons that will become clear in Section \ref{da}.

There is actually a rewriting of the relations above 
to the more familiar Gallavotti-Cohen type\footnote{One 
uses this term refering to fluctuation relations which can be given as 
the logarithmic ratio of probabilities of opposite events. The possible connection with 
the asymptotic time limit will be explained shortly.} fluctuation symmetries.

From \eqref{gc} by taking the function $f(x)= \delta(S(x)  - \sigma)$ we get
\begin{equation}\label{gc1}
\text{Prob}[S(X) = -\sigma] = e^{-\sigma}\,\text{Prob}[S(X) =\sigma]
\end{equation}
and, from \eqref{un} by choosing $f(x)=\delta(R(x) - r)$,
\begin{equation}\label{un1}
\text{Prob}[R(X) = -r] = e^{-r}\,\text{Prob}[R(X) =r]
\end{equation}
with probabilities referring to the probability law $P$.

Relations \eqref{un}--\eqref{afs} are general and can be applied in a variety of ways. 
A new application will be established in Section \ref{rem} as well as in
equations \eqref{gff}--\eqref{mira}. Moreover, from \eqref{gc} it follows that for all 
functions $g$ on $X$,
\begin{equation}
\langle g(X)\rangle - \langle g(\Theta X)\rangle = \langle g(X)\, S(X) \rangle _o
\label{eq:GenKubo}
\end{equation}
to first order in the action $A$ of \eqref{act} and where the last expectation $\langle \cdot \rangle _o$ 
is with respect to $P_o$.  
The same holds replacing $\Theta\rightarrow \Theta\Gamma$ 
and $S\rightarrow R$ in which case we would obtain fluctuation--activity relations.

Another consequence is that always 
$\langle R(X)\rangle \geq 0, \langle {\mathcal T}\Gamma - {\mathcal T}\rangle \geq 0$ and 
$\langle S(X)\rangle \geq 0$.  These inequalities are not only useful to determine the direction of currents 
(the more standard application) but, as we will see, enable to predict at what side 
of a boundary driven system the activity is 
largest, which is an entirely new application.

Let us emphasize that the results above are exact, i.e., valid for all times.
All the consequences mentioned 
will remain basically intact also for variables that 
differ from $S$ or $R$ by a total (time-)difference as long as some boundedness 
of that difference can be assured.
If so, we will get asymptotic fluctuation symmetries, where \eqref{gc}--\eqref{un1} 
are not exact for the variables but only valid in some limit (of large observation time). 
Such asymptotic formul{\ae} would correspond to {\it stationary} fluctuation theorems as in \cite{gc}. \\ 

Further, the relevance of the fluctuation identities \eqref{gc}--\eqref{un} 
depends crucially on the systematic and operational meaning of $S$ and $\mathcal T$.  
It was understood before that $S$ is deeply related to changes in entropy 
(as we will briefly repeat in the next Section); in Sections 
\ref{da}--\ref{slga} we treat a number of examples where 
$\mathcal T$ is made visible 
and related to the dynamical activity.

\section{Standard example: entropy flux}\label{se}
The present section contains the standard application of \eqref{gc} 
to obtain a fluctuation symmetry for the total entropy flux 
in nonequilibrium Markov jump models. The reader will only find as new 
some reflections towards the end of the section connecting the fluctuation 
symmetry also with response theory and the Kubo formula.
Nevertheless examples for spatially extended systems are not so common 
in the literature on fluctuation symmetries 
and the present section treats them in a still less familiar but unifying framework.

Consider a Markov jump process on a finite state space $K$.  
We specify the transition rates $k_t(x,y)$ 
(time-dependent) for jumps $x\rightarrow y$ between system states
\begin{equation}\label{dyn}
k_t(x,y) = \psi(x,y)\,\exp \{\frac{\beta_t}{2}[U(x,a_t) - U(y,a_t) + F(x,y)]\}
\end{equation}
where $a_t$ is a time-dependent (external) protocol changing the function $U$.  
$U(x,a)$ is called the energy of the system when it is found at state $x$ with 
external value $a$; this is because we imagine that 
the changes in $U$ are exactly balanced by the change of energy in the environment.
The driving $F(x,y)=-F(y,x)$ is antisymmetric but it does not need to be a total difference for all $x\rightarrow y$,
which is important to model nonequilibrium features. 
The reactivities $\psi(x,y)=\psi(y,x)$ are symmetric.   
The additional time-dependent parameter $\beta_t\geq 0$ in \eqref{dyn} is 
the varying inverse temperature of the environment (in units where $k_B=1$). 
Note that the nonequilibrium 
driving sits entirely in 
the function $F$ (which contains the irreversible work) and in the time-dependence of 
both the protocol $a_t$ and the inverse temperature $\beta_t$.
Clearly, if $F=0$ and 
when $a_t=a, \beta_t=\beta$ are constant, then the process is reversible 
with stationary distribution $\rho^\beta(x) \propto \exp -\beta U(x,a)$. 

It is important to note that both the 
form and the interpretation of \eqref{dyn} follow from 
the condition of local detailed balance, 
for which at all times $t$ (and always with $k_B=1$)
\begin{equation}
\log \frac{k_t(x,y)}{k_t(y,x)} = S_t(x,y)
\end{equation}
is the entropy flux in the transition $x\rightarrow y$; that is the change of entropy in the environment. This explains the standard 
origin of the exponential form of the rates $k(x,y)$ and in particular why the antisymmetric term $F(x,y)$ 
contributes to the irreversible work; see also below in \eqref{sout} and in the examples of Section \ref{kawad}.  

For a given path $X = (x_t, t\in [0,T])$ over the time-interval $[0,T]$ the energy change 
is given by
\begin{equation}
U(x_T,a_T) - U(x_0,a_0) = 
\sum_{t\leq T}\,\big[U(x_{t},a_t)- U(x_{t^-},a_t)\big] + 
\int_{0}^{T} \frac{\partial U}{\partial a_t}(x_t,a_t)\,\dot{a}_t \,\id t \label{1st}
\end{equation}
where the sum is made for the transitions $x_{t^-}\rightarrow x_t$ occurring 
at the jump times $t$, 
and $x_{t^-}$ denotes the state of the system right before the jump to $x_t$. 

In equation \refeq{1st} we have two effects for the energy change. Firstly, 
for fixed value $a_t$ the system 
state has changed and then energy is exchanged with the environment as heat
\begin{equation}\label{b1}
Q_o(X) := \sum_{t\leq T}\,\big[U(x_{t},a_t)- U(x_{t^-},a_t)\big]
\end{equation}
(again the sum is made over jump times in $X$).
Secondly, for fixed state $x_t$ the external value changes $\dot{a}_t = \frac{\id a_t}{\id t}$, doing work
\begin{equation}
W_o(X) := \int_{0}^{T} \frac{\partial U}{\partial a_t}(x_t,a_t)\, \dot{a}_t\,\id t.
\end{equation}
Thus, equation \eqref{1st} mimics the first law of thermodynamics. The energy change of the system equals 
the change in internal energy 
received as heat $Q_o$ from the environment plus the amount of work $W_o$ done on the system by the environment:
\begin{equation} U(x_T,a_T)-U(x_0,a_0) = Q_o(X) + W_o(X). \end{equation}
The nonequilibrium driving $F$ can be added and subtracted from this 
balance. We think of it as doing work on the system, 
which is instantaneously released as heat, so that now
$U(x_T,a_T) - U(x_0,a_0) = Q(X) + W(X)$, but with
\begin{equation}
\label{b2}
Q(X) := Q_o(X) - \sum_t F(x_{t^-},x_t),\quad W(X) := W_o(X) + \sum_t F(x_{t^-},x_t)
\end{equation}
with all terms depending on a specific path $X$.  We refer to \cite{seki} for more details and insights 
on stochastic energetics.

In the same spirit we can also associate a change in entropy of the environment to a path or 
trajectory $X$.  The idea is that the environment 
consists of big equilibrium reservoirs undergoing only reversible changes in interaction with the system. 
One looks back at \eqref{b1} and \eqref{b2} to define
\begin{equation}\label{sout}
S_{\out}(X) := -\sum_t\beta_t \delta Q^{(t)}= \sum_{t}\beta_t\,\{F(x_{t^-},x_t) -\big[U(x_{t},a_t)- U(x_{t^-},a_t)\big]\}
\end{equation}
for the change of the entropy in the environment (always per $k_B$). This term 
is the total entropy flux for trajectory $X$, which can be split into a reversible part, 
due to the energy exchange, and an irreversible part
\begin{equation}\label{irr}
\sigma(X) := \sum_s \beta_s \,F(x_{s^-},x_s).
\end{equation}
Particular examples such as the one treated in Section \ref{kawad} 
will present explicit expressions for the driving $F$, as 
in equations \eqref{rbdrivekw}--\eqref{bdrivekw}. 
The examples treated will also clarify further 
its connection to irreversibility, which as will be seen 
in the models sits exclusively at the boundaries.

We now repeat the observation of \cite{maes,time} that the entropy flux \eqref{sout} can be obtained as for \eqref{gc} 
as the source term of time-reversal breaking.\\

Let us leave out the kinematical time-reversal $\pi$ on $K$ and proceed with the undecorated 
time-reversal $\Theta$, which is defined on trajectories in phase space or paths $X$ 
via $(\Theta X)_t = X_{T-t}$ for $t\in [0,T]$.  
One can check from \eqref{sout} that the entropy flux per path is antisymmetric under time-reversal, 
$S_{\out}(X) = -S_{\out}(\Theta X)$.  
Let now $P_\mu$ denote the path distribution when we start at time zero from a probability law $\mu$ on $K$.
The time-dependence of the protocol 
can be reversed to define
$\tilde{k}_t(x,y) := k_{T-t}(x,y)$.  We choose a second probability law $\nu$ on $K$ for 
starting the latter (protocol-reversed) Markov process, with path-distribution denoted by $\tilde{P}_\nu$. 
Assuming $\mu,\nu>0$ 
and  that $k_t(x,y)=0$ implies  $k_t(y,x)=0$ (dynamical reversibility),  
we can find the $S$ in \eqref{gc} via
\begin{equation}\label{sour}
\frac{\id P_\mu}{\id \tilde{P}_\nu\Theta} = e^{S} 
\end{equation}
and find
\begin{equation}
 S(X) = \log\frac{\mu(x_0)}{\nu(x_T)} + \log \frac{k_{t_1}(x_0,x_{t_1}) k_{t_2}(x_{t_1},x_{t_2})
\ldots k_{t_n}(x_{t_{n-1}},x_{T})}{k_{t_n}(x_T,x_{t_{n-1}})\ldots k_{t_2}(x_{t_2},x_{t_1})k_{t_1}(x_{t_1},x_0)}
\label{srevs}
\end{equation}
for jump times $t_1, t_2,\ldots, t_n$ in $X$.  Indeed the jump times in the reversed trajectory 
$\Theta X$ are respectively 
$T-t_n,\ldots,T-t_2,T-t_1$. One can see what \eqref{srevs} becomes for the rates \eqref{dyn}.
Substituting into the previous formula makes
\begin{equation}\label{rrr}
S(X) - \log\frac{\mu(x_0)}{\nu(x_T)} = \sum_t \beta_t \{U(x_{t^-},a_t) - U(x_t,a_t) + F(x_{t^-},x_t)\}
\end{equation}
which is \eqref{sout}.  That relation can be called a (generalized) Crooks relation \cite{crooks}, 
and for $F\equiv 0$ it almost immediately 
produces Jarzynski identities 
which are used to evaluate equilibrium free energies from the fluctuations of the dissipative work --- we refer to 
the literature and the references therein 
for more details, \cite{jarz,rito,poincare}.

Let us now specify to the case where $\beta_t=\beta, a_t = a$ are constant in time. In particular, 
with respect to \eqref{irr}, 
and for state functions $h_\mu(x) := \log \mu(x) + \beta U(x) , h_\nu(x) := \log \nu(x) + \beta U(x)$, 
we have the identity   
\begin{equation}\label{fls}
S(X) = \beta\sum_t F(x_{t^-},x_t) + h_\mu(x_0) - h_\nu(x_T)
\end{equation}
for all trajectories $X$.
Note that the left-hand side is defined from \eqref{sour} implementing \eqref{gc}, while the right-hand side is 
defined from the heat and \eqref{sout}--\eqref{irr}. Therefore, the identities \eqref{rrr}--\eqref{fls} 
are the core of what is 
generally called the fluctuation symmetry, the fluctuation relations or the fluctuation theorem 
(transient or steady state, as recalled also at the end of Section \ref{gobs}) for the entropy production.  

\subsection{Exact fluctuation symmetry}

In the following we restrict ourselves to time-homogeneous Markov processes and we do no longer write the 
dependence on $a_t=a$.  We take also inverse temperature $\beta=1$.

Consider the reference reversible process $P_o$ started in equilibrium $\rho_o$ for which there is 
detailed balance with rates
  \begin{equation}
  k_o(x,y) = \psi(x,y)\,e^{\frac 1{2}[U(x)-U(y)]},\quad \rho_o(x) =\frac 1{Z} e^{-U(x)}.
  \end{equation}
The nonequilibrium process has rates $k(x,y) =  k_o(x,y)\,\exp F(x,y)/2$ 
and we choose to start it also from $\rho_o$. 
Its distribution on paths $X$ in the time-interval $[0,T]$ is then denoted by $P$. 
We proceed as in \eqref{defs} to find
\begin{equation}\label{star}
S(X) = \sum_t F(x_{t^-},x_t), \quad \mathcal{T}(X) =
2\int_0^T[\xi(x_s) - \xi_o(x_s)]\id s
\end{equation}
for 
escape rates $\xi(x) := \sum_y k(x,y)$.  
Now clearly \eqref{gc} holds, and with $f(X) = \exp[-z S(X)]$ for all $z\in \bbC$, 
we have the exact fluctuation symmetry
\begin{equation}\label{exa}
\langle e^{-z S(X)} \rangle= \langle e^{-(1-z) S(X)} \rangle
\end{equation}
with expectations in the nonequilibrium process starting from the equilibrium 
distribution $\rho_o$. That result in itself is of course not new and has been 
derived in various ways; see e.g. equation (2.32) in \cite{maes} or equation (3.34) in \cite{har07}.

Another way to get an exact fluctuation symmetry is to look back at \eqref{fls} with probabilities 
$\nu=\mu=\rho$ equal to 
the stationary distribution 
of the nonequilibrium process. 
We then have by combining \eqref{sour} with \eqref{fls} that in the nonequilibrium 
steady regime, for all $T$,
\begin{equation}\label{ffss}
\langle f(X)\rangle = \langle e^{-\sigma(X) - h(x_0) + h(x_T)} f(\Theta X)\rangle
\end{equation}
for irreversible entropy flux $\sigma(X) = \beta\sum_t F(x_{t^-},x_t)$
and with state function $h(x) :=  \log \rho(x) + \beta U(x)$. 
The exact symmetry \eqref{ffss} would invite us 
to give special physical meaning also to that function $h$, 
but no convincing 
thermodynamic or operational meaning exists. 
Only in some cases 
like the models we treat in Sections \ref{bd0rangep}--\ref{slga}, 
this physical interpretation of $h$ in \refeq{ffss} 
can be made. 
This is also why asymptotic (in $T\uparrow +\infty$) fluctuation symmetries, 
obtained from \eqref{ffss} for $f$ any positive function of $\sigma(X)$, 
have been more appreciated. These asymptotic fluctuation formulas are obtained 
by taking the logarithm of both sides in \refeq{ffss} and dividing them by $T$; then 
using the boundedness of the function $h$ will make it disappear when finally 
letting $T\uparrow +\infty$.

\subsection{Relation to linear response}\label{rem}

Looking backward, it appears that the main input has been relation \eqref{rrr}. 
That has analogues for diffusion process \cite{LS,kurch,jmp}, for dynamical systems \cite{gc,bon,ruelle,verb} 
and also for non-Markovian processes \cite{maes,an,semim} 
as long as there is sufficient space-time locality to ensure a large deviation principle \cite{maes}. 
The main origin of the fluctuation symmetry is therefore the identification of the entropy flux as marker 
of time-reversal breaking, \cite{bon,maes,crooks,time}.

Quite some features of the close-to-equilibrium regime are easily deduced from the fluctuation symmetry. 
There are for example 
the Green-Kubo relations, 
with Onsager reciprocity as first explained in \cite{gal} following from an extended fluctuation symmetry. 
More globally, the validity of 
the McLennan ensemble close-to-equilibrium is another implication, 
see \cite{koma,mcl}.

We illustrate just one aspect which we have not seen stated as such, and which is useful.
Start again from \eqref{gc} and take a function $f(X) = g(\Theta X) - g(X)$ in terms of another function $g$ of interest. 
Then,
\begin{equation}\label{gf}
\langle g(X)\rangle = \langle g(\Theta X)\rangle + \langle (g(\Theta X) -g(X))  \,e^{-S(X)}\rangle
\end{equation}
Imagine now that the action $A$ in \eqref{act} is small, so that the law $P$ is just a small perturbation 
of the reference law $P_o$ and so that $S = A\Theta - A$ is small.
We can then expand the last term in \eqref{gf} to bring
\begin{eqnarray}\label{gff}
\langle g(X)\rangle &=& \langle g(\Theta X)\rangle 
+ \langle g(\Theta X) -g(X)\rangle - \langle (g(\Theta X) -g(X)) S(X)\rangle_o\nonumber\\
&=& \langle g(\Theta X)\rangle + \langle g(X)\, S(X)\rangle_o
\end{eqnarray}
where the last expectation, with the subscript $\langle \cdot\rangle_o$, is with respect to the reference $P_o$ and we have 
used that $P_o$ is $\Theta-$invariant.  That linear order relation 
can be applied to the context of dynamical ensembles as we had it above, 
with $\Theta$ being time-reversal on trajectories $X=(x_t, t\in [0,T])$. 
Take for example the particular case where 
$g(X)= O(x_T)$ so that $g(\Theta X) = O(x_0)$ for a state function $O$; 
$x_0,x_T$ are the initial and final states of the trajectory $X$, respectively.
We then obtain from \eqref{gff} the linear response formula
\begin{equation}\label{kub}
\langle O(x_T)\rangle = 
\langle O(x_0)\rangle + \langle O(x_T)\,S(X)\rangle_o
\end{equation}
where the expectations refer to the process $P$ started from equilibrium $\rho_o$ at time zero. In order to 
recognize the Kubo formula one should 
substitute in \eqref{kub} the expression \eqref{rrr} for $S(X)$ 
with $F\equiv 0$, $\beta_t\equiv \beta, a_t= a - \varepsilon_t \theta(t)$ and $\mu=\nu=\rho_o$ being the equilibrium 
distribution with potential $U(x,a)$. 
Then, still using the first law \eqref{1st}, we arrive at the more familiar Kubo expression
\begin{eqnarray}\label{mira}
\langle O(x_T)\rangle - 
\langle O(x_0)\rangle_o &=&\\
\langle O(x_T)\,S(X)\rangle_o 
&=& \int_0^T\id s \,\varepsilon_s \frac{\id}{\id s}\langle O(x_t) \frac{\partial}{\partial a}U(x_s,a)\rangle_o\nonumber
\end{eqnarray}
Yet, it takes the combination \eqref{rrr}--\eqref{kub} to immediately understand why this formula 
is truthfully called fluctuation-{\it dissipation} relation.\\

Moving beyond the linear response around equilibrium makes it more difficult to 
find specific consequences.  Of course, 
the fluctuation relations hold unperturbed but there is no direct way to derive 
more specific results.  In fact, it appears 
that one really needs more information about the time-symmetric part, 
$\mathcal{T}$ in \eqref{defs}, to move further \cite{mat,beyond}; 
that is also part of the motivation of the next sections.

\section{Symmetry in dynamical activity}\label{da}

We come to give examples of the fluctuation symmetry \eqref{un}, 
referred to in the title of the paper as {\it active} because they 
deal with the dynamical activity.

\subsection{Boundary driven Kawasaki dynamics}
\label{kawad}
We consider a system of indistinguishable particles subject to exclusion on a lattice interval 
which is boundary driven. 
The state space is $K = \{0,1\}^{\{1,2,\ldots,L\}}$, where 
states are particle configurations $x=(x(i),i\in \{1,2\ldots,L\}), x(i)=0,1$, 
interpreted as vacant {\it versus} occupied sites on a lattice interval. 
The dynamics has two parts. First, there is a bulk exchange of neighboring occupations with rate given, 
for inverse temperature $\beta\geq 0$,
\begin{equation}
k(x,y) = \exp -\frac{\beta}{2}[V(y)-V(x)]
\end{equation}
when $y(j)=x(j)$ for all $j$ except for $y(i)=x(i+1), y(i+1)=x(i)$ for some $i=1,2\ldots,L-1$. The 
interaction between neighboring sites is ruled by the potential
\begin{equation}
U(x)=-\kappa\sum_{i=1}^{L-1}x(i)x(i+1)
\end{equation}
where $\kappa \in \bbR$ is the coupling parameter; note that the case $\kappa =0$ 
corresponds to the 
symmetric exclusion process. 

Second, apart from the 
interacting diffusion part to the dynamics above, 
there are also the reactions at the boundary sites where 
creation and annihilation of particles take place 
\begin{equation}\label{rbdrivekw}
k(x,y) = \exp -\frac{\beta}{2}[U(y)-U(x)]\,\exp\frac{\beta}{2} F(x,y)
\end{equation}
for $y(j)=x(j)$ except for $j=i$, the boundaries, where $y(i)=1-x(i)$ 
with $i=1$ and $i=L$. 
Besides
\begin{eqnarray}\label{bdrivekw}
F(x,y) &=& +(a+c_i\delta )\mbox{ when } y(i)=1,x(i)=0, \,y(j)=x(j), j\neq i,\nonumber\\
&=& -(a+c_i\delta )\mbox{ when } y(i)=0,x(i)=1,\, y(j)=x(j), j\neq i
\end{eqnarray}
where for $i=1,L$ one has $c_1=1, c_L=-1$ and for some fixed parameters 
$a,\delta \in \bbR$.  The physical interpretation of this birth and death process 
is the contact  at the boundaries with particle reservoirs at left and right chemical potentials 
$\mu_1  := a + \delta, \mu_L := a - \delta $, respectively. 

For all other transitions we have $k(x,y) = 0$.  As a result,
\begin{equation}
k(x,y) = k_o(x,y)\,\exp[\frac{\delta \beta}{2}\,J(x,y)]
\end{equation}
with $k_o(x,y) = \exp [{\cal S}(y)-{\cal S}(x)]/2, {\cal S}(x) := -\beta U(x) + a\beta {\cal N}(x), 
{\cal N}(x) := \sum_{i=1}^L x(i)$ 
(number of particles in the system for state $x$), and with {\it current}
\begin{eqnarray}
J(x,y) &=& +c_i \mbox{ when a particle enters at } i=1,L\nonumber\\
&=& -c_i \mbox{ when a particle leaves }
\end{eqnarray}
and zero otherwise.
In other words, $J(x,y) = J_L(x,y) - J_1(x,y)$ with $J_1(x,y)$ 
the current of particles into the left reservoir 
and $J_L(x,y)$ the current of particles into the right reservoir 
for the transition $x\rightarrow y$.

For $\delta=0$ (and only for $\delta=0$) there is detailed balance with grand-canonical ensemble
\begin{equation}
\rho_o(x) = \frac 1{\cal Z}\,\exp {\cal S}(x).
\end{equation}
In this case the parameter  $a$ is the chemical potential of both particle reservoirs left and right. 
That equilibrium process determines our reference distribution $P_o$.
Nonequilibrium arises from taking 
$\delta \neq 0$, which makes the chemical potentials in the imagined left and right particle reservoirs different. 
We can start the nonequilibrium process from the same $\rho_o$, giving our distribution $P$, 
but asymptotically in time a nonequilibrium steady regime will develop. 
In particular it is easy to prove now that for $\delta>0$ there will be a steady particle current from left to right. 
See for example \cite{prague} for the details of the standard fluctuation symmetry as in the previous section.

The decomposition \eqref{defs} here gives 
\begin{equation}\label{sx}
S(X) =  
\beta \delta \left[ J_1 (X )- J_L(X )\right]
\end{equation}
with $S(\Theta X) = -S(X)$ for $\Theta$ time-reversal, and $J_1(X) := \sum_t J_1(x_{t^-},x_t)$ 
the net number of particles 
that escape from the lattice interval to the left particle reservoir.
Note that $ J_L(X) + J_1(X) = -{\cal N}(x_{T}) + {\cal N}(x_0) $, 
the change of the number of particles in the system.

For the time-symmetric part of the action we can compute, from \eqref{star} 
\begin{equation}\label{dyna}
\mathcal{T}(X) = 2\int _0^{T}\id t\left[ B_1(x_t;a,\delta ) + B_L(x_t; a, \delta )\right]
\end{equation}
where (putting now $\beta=1$ for notational simplicity)
\begin{eqnarray}
B_{i}(x;a,\delta ) & :=& e^{(a+ c_i\delta)/2} - e^{a/2}  + \{e^{-(a+c_i\delta)/2}
-e^{(a+c_i\delta)/2} + e^{a/2} - e^{-a/2}\}x(i)\nonumber\\
&+& (e^{(a+c_i\delta)/2} - e^{a/2})(e^{\kappa/2}-1)\,x(i+c_i)\nonumber\\
&+& \{(e^{-\kappa/2}-1)(e^{-(a + c_i \delta)/2} - e^{-a/2})\nonumber\\
&-& (e^{\kappa/2}-1)(e^{(a+c_i\delta)/2} - e^{a/2})\}x(i)x(i+c_i)\nonumber
\end{eqnarray}
again for $i=1,L$ and $c_1=1, c_L=-1$.
Next, in order to obtain the symmetry in the dynamical activity, 
we apply the mirror transformation $\Gamma$ through which $(\Gamma X)_t(i) = X_t(L-i+1)$. 
Observe that in that mirror symmetry $ J_1 (X ) =  J_L (\Gamma X ), S\Gamma (X)= -S(X)$.  
We can thus compute 
\begin{equation}
R (X) =\frac{1}{2}\left( {\mathcal T}(\Gamma X)
-{\mathcal T}(X) \right) =\int_0^T\id t \,r(x_t) \label{Rvariable}
\end{equation}
from the expected difference in transitions (jumps in and out of the system) left {\it versus} right, 
to find
\begin{eqnarray}
r(x) &=& \sum _{i=1,L} \{ 
2\sinh\frac{\delta}{2}\;\big((e^{-\kappa/2}-1)e^{-a/2}  + (e^{\kappa/2}-1)e^{a/2}\big) 
c_ix(i)x(i+c_i)\nonumber\\
&-& 2\big( \sinh\frac{a-\delta}{2} - \sinh\frac{a+\delta}{2} \big)c_ix(i)\nonumber\\ 
&-& 2e^{a/2}\,\sinh\frac{\delta}{2}\,(e^{\kappa/2}-1)\,c_ix(i+c_i)\}\label{R}
\end{eqnarray}
which is of course also odd in the driving field $\delta$. Now 
for the boundary driven symmetric exclusion process we must take the coupling $\kappa=0$, 
and in \refeq{R} only survive 
\begin{equation}\label{ko}
r^{\kappa=0} (x)= 
 2\big( \sinh\frac{a-\delta}{2} - \sinh\frac{a+\delta}{2} \big)(x(L)-x(1)),
\end{equation}
which is given entirely in terms of the difference in occupations at the outer sites.

It follows from the general analysis in Section \ref{gobs} that $R(X)$ in \eqref{Rvariable} 
verifies the fluctuation symmetries \eqref{un}--\eqref{un1}. This is a non-trivial general 
identity whose meaning refers to the reflection-antisymmetric part in the dynamical activity \eqref{dyna}. 
In particular, that identity \eqref{un} for that same $R$ in \eqref{Rvariable}--\eqref{R} remains strictly 
valid even when modifying the interaction potential $U$ in the bulk of the system.
On the other hand, applying the general consequence that 
$\langle R(X)\rangle \geq 0$, or $\sum_x r(x)\,\rho(x) \geq 0$, to \eqref{ko} only gives the well 
known fact that the density is larger (for constant temperature) 
at the side of the largest chemical potential.

\subsection{Boundary driven zero range process}
\label{bd0rangep}
We now discuss the application of fluctuation symmetries to a bosonic version of the previous example, 
where particles diffuse 
without exclusion principle.

Consider again a one-dimensional channel composed of $L$ cells in which we observe 
occupation numbers $n(k)\in \bbN$, $k=1,\dots ,L$. 
The particle configuration $x=(n(1),\ldots,n(L))$ can change in two ways. 
In the first place, it changes at a rate $w(n(i))$ via bulk hopping, 
$x\rightarrow x-e_i+e_{i\pm 1}$, where $e_i$ stands for the particle 
configuration with one particle in cell $i$ and zero elsewhere. 
The choice $w(n(i)) \propto n(i)$ corresponds to independent particles. 
Secondly, at the boundaries, the channel is connected to left/right particle 
reservoirs with chemical potentials 
$\mu_{1}$ and $\mu_L $, respectively. 
The transition rates for the creation/annihilation of particles at the two sites $i=1,L$ are then
\begin{eqnarray}
k(x,x-e_i) &=& s_i \,w(n_i) \nonumber \\
k(x,x+e_i) &=& r_i\,e^{c_i\delta}
\end{eqnarray}
with $c_1=1,c_L=-1$.
The rates for these transitions evoke the chemical 
potentials at the boundary walls 
from $\mu _i=\log \left(r_i/s_i\right)+c_i\delta $. 
We assume that $s_1/r_1= s_L/r_L$ so that, for $\delta=0$, we have 
the equilibrium situation where the chemical potentials left and right become equal. 
Of course we could have chosen also to modify the exit rates $s_i$ but it appears physically 
most accessible to change the incoming rates $r_i\rightarrow r_ie^{c_i\delta}$ 
to achieve a nonequilibrium regime, as we also do in the next section.  
In fact, to make the equilibrium left/right symmetric we also take
$s_1=s_L, r_1=r_L$. 
The corresponding stationary distributions $\rho_o$ (at $\delta=0$) and $\rho$ (at general $\delta$) are 
product distributions but that will not be used in the following.

Consider the trajectories $X = (x_t, t\in [0,T])$. 
Both equilibrium $P_o$ and nonequilibrium $P$ processes 
start from the same equilibrium distribution $\rho_o$. 
The action \eqref{act} is  easily calculated to be 
\begin{equation}
A(X)= \delta \left( I_{1}^{\shortrightarrow }(X) - I_{L}^{\shortleftarrow }(X) \right) + 
T \left[ \left( r_1 + r_L \right) \left( e^{\delta } -1 \right) \right]
\end{equation}
where e.g., $I_{1}^{\shortrightarrow }(X)$ indicates the number of particles entering the 
system from the left reservoir for the path $X$.
As we apply time-reversal $\Theta$, we obtain the time anti-symmetric part 
of the action $ S(X)= A(\Theta X) - A(X)$ 
\begin{eqnarray}
S &=& \delta \left[ \left( I_{1}^{\shortleftarrow } - I_{1}^{\shortrightarrow }\right) +
 \left( I_{L}^{\shortrightarrow } - I_{L}^{\shortleftarrow }\right) \right] \nonumber \\
&& =\delta(J_1-J_L)
\end{eqnarray}
where now e.g., $J_1 := I_{1}^{\shortleftarrow } - I_{1}^{\shortrightarrow }$ is the net number 
of particles that have escaped 
to the left particle reservoir during $[0,T]$. 
As usual and as explained before, that entropy production satisfies the exact fluctuation symmetry \eqref{gc1}.  
For the asymptotic form, one must be more careful because of the unbounded number of particles; see \cite{rose}. 
Here we are however more interested in the dynamical activity.\\

Let us then look at the time-symmetric term $ \mathcal{T}(X)= A(\Theta X) + A(X) $, 
\begin{equation}\label{but}
\mathcal{T} = 
\delta\,\left[ \left( I_{L}^{\shortrightarrow } + I_{L}^{\shortleftarrow }\right) -
\left( I_{1}^{\shortrightarrow } 
+ I_{1}^{\shortleftarrow }\right) \right]
-2\left[ \left( r_1 + r_L \right) \left( e^{\delta } -1 \right) \right]\;T.
\end{equation}
This is the analogue to \eqref{dyna}, for the Kawasaki dynamics example.
Thus, as we did in Section \ref{kawad}, we will apply the mirror transformation $\Gamma$, 
reversing left/right. 
First note that here again the entropy $S$ is antisymmetric under $\Gamma$, $S\Gamma = -S$. 
On the other hand, we have
\begin{equation}
\mathcal{T}(\Gamma X) - \mathcal{T}(X) 
=2\delta\,\left(I_{1}^{\shortrightarrow } + I_{1}^{\shortleftarrow} 
- I_{L}^{\shortrightarrow } - I_{L}^{\shortleftarrow }\right) 
\label{tsymvar}
\end{equation}
exactly proportional to the difference in dynamical activity 
between the right and left boundaries, 
\begin{equation}
\label{actvdiff}
\Delta(X) := I_{L}^{\shortrightarrow } + I_{L}^{\shortleftarrow} 
- I_{1}^{\shortrightarrow } - I_{1}^{\shortleftarrow }
\end{equation}
Following the logic of \eqref{un}, that suffices for a variable 
$\mathcal{T}\Gamma - \mathcal{T}\propto \Delta $ 
to satisfy a fluctuation symmetry \eqref{un1} up to a total time-difference.
Even more, when the observable $f\Theta = f$ is time-symmetric, then 
\begin{equation}\label{zrf}
\langle f(\Gamma X)\rangle = \langle f(X) \,e^{\delta\,\Delta(X)}
\rangle
\end{equation}
for all times $T$, where we start the nonequilibrium process at time zero from $\rho_o$.  
For example taking $f=\Delta$, to first order in $\delta$, 
\begin{equation}
\langle \Delta(X) \rangle = -\frac{\delta}{2}\,\langle \Delta^2(X) \rangle_0^{\text eq}\label{poe}
\end{equation}
which is formally similar to a Green-Kubo relation \cite{gal,kubo} but now 
the observable $\Delta$ in \refeq{actvdiff} is time-symmetric.

It is in fact true for all $\delta\geq 0$ that $\langle \Delta\rangle \leq 0$, which means that the 
greatest activity is to be found at the boundary side of the largest chemical potential.  
In other words, as for the boundary driven Kawasaki dynamics also for zero range, 
the particle current can be said to be directed away from the region of largest activity. 
These statements all hold for any form of the bulk rate $w$ and are quite independent of 
the usual statements involving the fluctuation symmetry of entropy production or currents.

\section{Spinning Lorentz Gas}\label{slga}

\begin{center}\begin{figure}\includegraphics[width=\textwidth]{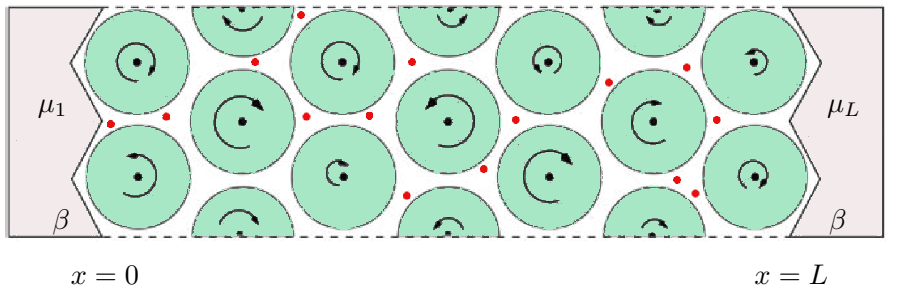}
\caption{In the spinning Lorentz gas  $M$ disks 
with radius one and centers fixed in a triangular lattice rotate freely and exchange energy 
with point particles (of mass one) via elastic collisions \cite{Lar03}. The particles evolve via classical 
mechanics inside the slab of length $L$ with
periodic boundary conditions in the vertical coordinate. The slab is placed 
among thermo-chemical reservoirs (ideal gases) with (for the present paper) 
equal inverse temperatures $\beta$ 
and different chemical potentials $\mu _{i=1,L}$. Particles 
can enter and leave to/from the reservoirs at the left and right boundaries. }
\label{fig:slg}\end{figure}\end{center}

The spinning Lorentz gas (SLG) is a classical mechanical model of particle scattering in 2D; 
it is actually an interacting version of the normal Lorentz gas \cite{Lor05}, which is 
a well known example of deterministic particle diffusion \cite{Gas98, Sza00}. The SLG has 
the additional feature of providing local thermalization of the wandering particles along 
with the scatterers; a complete description of this and the coupled energy and 
mass transport properties of the SLG model can be found in \cite{Lar03}. As a matter of fact, 
the validity of the fluctuation theorem for the entropy production (Eqn. \refeq{gc1}) and for
the joint distribution of currents has been tested for this model, of course taking 
into account the limitations due to the unbounded kinetic energy, see \cite{Sal09}.
Also, a precise meaning to the state function $h$, mentioned after \eqref{ffss}, can be found in 
the SLG model for the exact symmetry case, which is then taken to the asymptotic limit 
where $h$ vanishes \cite{Sal09}.\\
As illustrated in Fig.~\ref{fig:slg}, the array of scatterers is connected to 
thermo-chemical reservoirs, with chemical potentials $\mu _i, i=1,L$ and at inverse temperatures $\beta$. 
This setting drives the system into a nonequilibrium stationary regime when $\mu _1\neq \mu _L$.

The SLG is a microscopic mechanical model which we want to connect with the boundary 
driven zero range model of the previous section.
In order to do this, note first that only at the walls in Fig.~\ref{fig:slg} 
point particles can enter and leave the system. The rates at which new 
particles enter are related to the mean density $u$ 
of their reservoir as $\propto u/\sqrt{\beta}$, an effusion process; 
see also \cite{Reif65}.
We focus in a nonequilibrium setting of the SLG where there is 
a reservoir chemical potential difference, given by 
$\beta \Delta \mu = \beta (\mu _L - \mu _1)=\log \left( u_L/u_1 \right)$; hence, 
in the notation of the previous section we have $2\delta = - \Delta\mu$.\\

The hypothesis to be tested here is that identical fluctuation 
relations as \eqref{un}--\eqref{un1} hold for 
the dynamical activity in the SLG as we had it for the boundary driven zero range process 
before, particularly in the version \eqref{zrf}.  One therefore looks back at the expression \eqref{tsymvar}. 
More precisely, 
we look at the fluctuations of the time-symmetric variable
\begin{equation}
R = \frac{\beta \Delta \mu}{2T } \left( 
(I_{1}^{\shortrightarrow } +
I_{1}^{\shortleftarrow } )
-( I_{L}^{\shortrightarrow } +
I_{L}^{\shortleftarrow } ) 
\right)
\label{slgtsymvar}
\end{equation}

\begin{center}\begin{figure}
\includegraphics{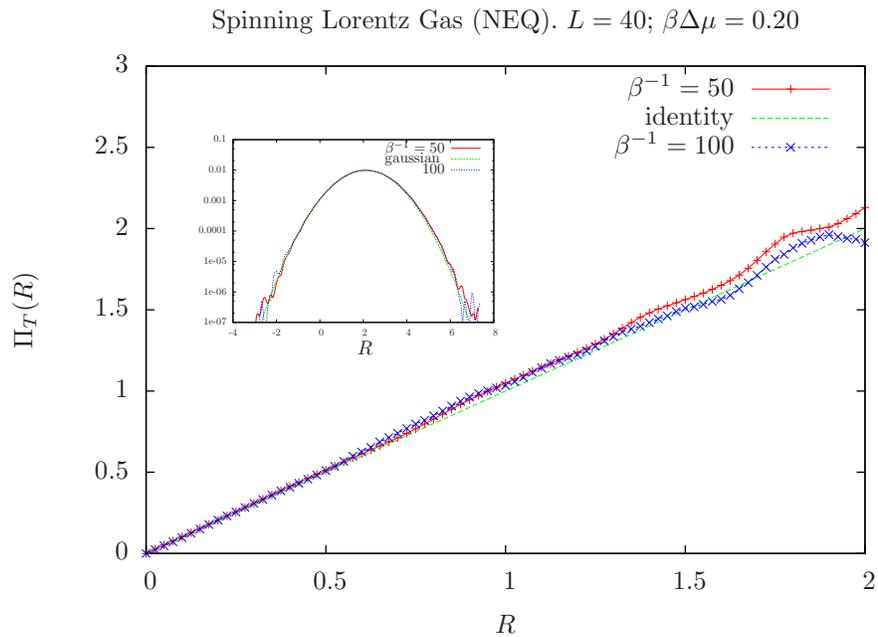}
\caption{The fluctuation symmetry for the difference in dynamical activity 
is tested numerically 
in nonequilibrium simulations of the SLG. In the inset, the 
probability distribution measured in the simulation $P_{T }\left( R \right)$ of $R$ 
in equation \refeq{slgtsymvar} is given.
The slab length is $L=40$, with reservoir chemical potential difference $\beta \Delta \mu=0.2$, and
reservoir temperatures $\beta^{-1}=50$ (crosses) and $\beta^{-1}=100$ 
(stars) giving identical result.
}
\label{figFTsymslg01}
\end{figure}\end{center}

\begin{center}\begin{figure}
\includegraphics{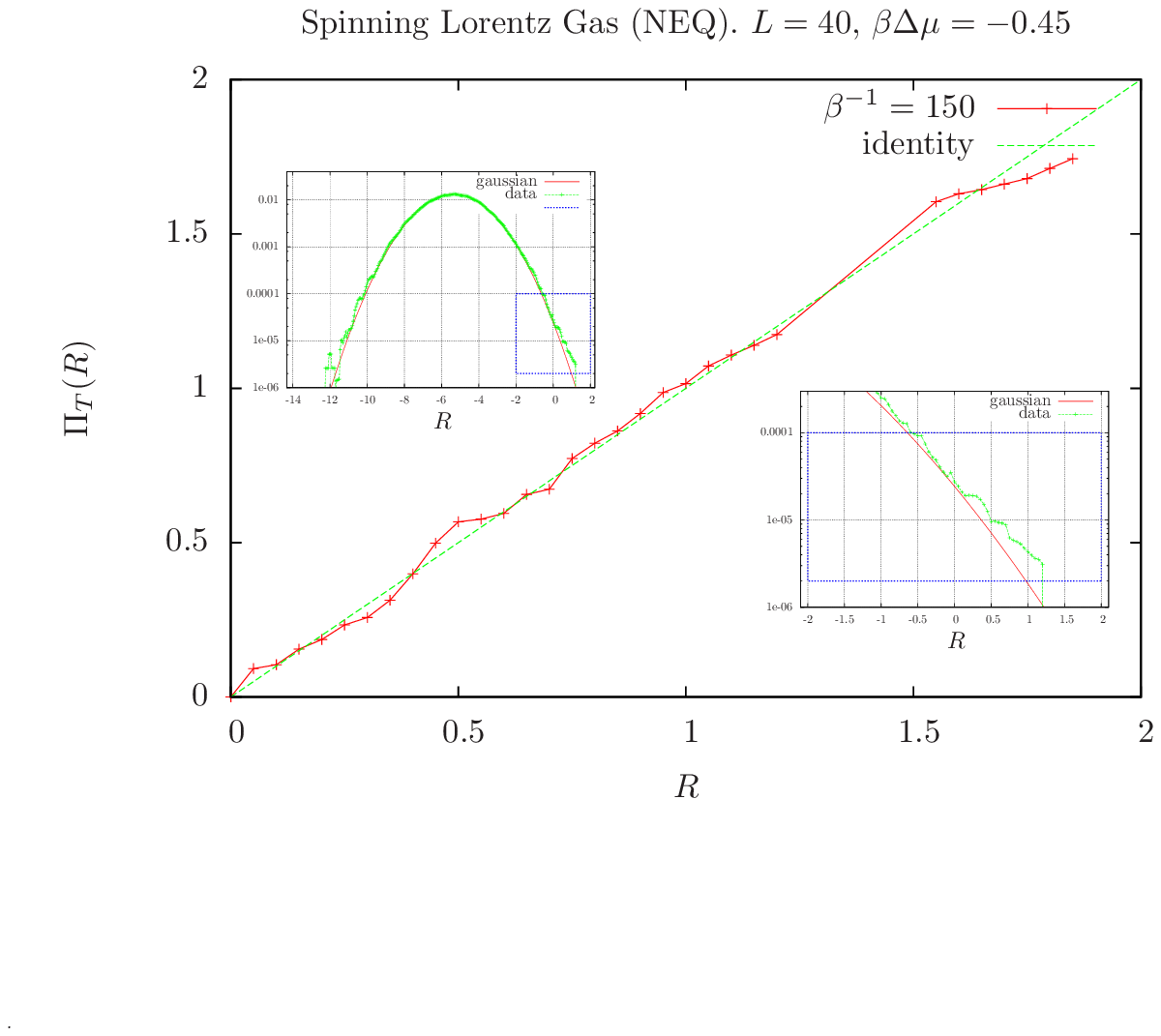}
\caption{ This figure shows the validation of the time-symmetric fluctuation 
theorem for the case of non-Gaussian fluctuations of the dynamical activity difference $R$, 
in the SLG in stationary nonequilibrium. The chemical potential difference $\beta \Delta \mu =-0.45$, 
$\beta =1/150$ and slab length $L=40$. The inset shows the probability $P_T(R)$ that was 
measured from the numerical simulation for a large measuring time $T = 4.0$. 
The interval of fluctuations 
around zero is far from the average value, where one distinguishes non-Gaussian behavior. 
In the main plot, the crosses show 
the evaluation of the fluctuation theorem for the probabilities in the inset; these 
data fit to a straight line with slope close to one, $m=0.99735\pm 0.01265$.
}
\label{figFTsymslg02}
\end{figure}\end{center}

We have measured in molecular dynamics simulations of the SLG model the 
probability distribution $P_{T}\left( R \right)$, in stationary nonequilibrium. 
Figs.~\ref{figFTsymslg01} and \ref{figFTsymslg02} show the validation of the time-symmetric fluctuation 
theorem for $P_{T}\left( R \right)$.
In these figures, to test the fluctuation symmetry
we plot, as usual, the functional
\begin{equation}
\Pi_T\left( R \right) = \frac{1}{T}\log\frac{P_T(R)}{P_T(-R)}
\label{ftfunct}
\end{equation}
The measuring time was a 
large value of $T = 4.0$; in the same time units, 
the average time between collisions in the gas is $\sim 2.5\times 10^{-3}$. 
In the first nonequilibrium case (Fig.~\ref{figFTsymslg01}) the stationary state 
is obtained by a chemical potential difference
$\beta \Delta \mu =0.20$ between the reservoirs, and for two different temperatures.
The second case (Fig.~\ref{figFTsymslg02}) corresponds to a 
larger driving $\beta \Delta \mu =-0.45$; this gives a fluctuation theorem 
interval in which the distribution is visibly non-Gaussian.\\ 

The variable \eqref{slgtsymvar} gives the fluctuations in the 
difference of dynamical activity at sites $i=1,L$. As in the remark around \eqref{ko}, 
here the dynamical activity in \eqref{slgtsymvar} is proportional to 
the number of transitions at each of the walls; in other words, it is 
proportional to the local boundary density. Since the temperature in this case 
is uniform, the activity fluctuations are simply related to density fluctuations 
of the stationary {\it profiles}. Thus, when measuring the differences in dynamical activity in 
\eqref{slgtsymvar} one obtains asymmetric statistics arising from the density profile in the slab, 
which is shaped by the nonequilibrium condition set by the reservoirs. 

\section{Summary}
We have discussed a general 
framework to derive Gallavotti-Cohen type fluctuation relations 
based on symmetry transformations applied to the dynamical ensemble.
We have shown that the time-antisymmetric sector 
contains the more usual fluctuation relations 
for the entropy production, with 
time reversal as the fundamental symmetry. 
In the other hand, fluctuation symmetries for time-symmetric variables
involve a different phenomenology, 
dealing with non-dissipative variables. 
The present paper indeed emphasizes the relevance of this less studied and
complementary time-symmetric fluctuation sector of the non equilibrium process.  
For this an extra symmetry is involved, most simply a mirror or reflection symmetry, 
which basically is equivalent to reversing the driving field. 
This task leads to fluctuation symmetry relations for 
differences in the dynamical activity, as we have illustrated 
with three examples of boundary driven systems.
Interestingly, one finds that new Green-Kubo relations for the activity hold, 
and we understand where in some spatially extended system the activity is maximal.

The fact that the same time-symmetric fluctuation symmetry 
remains verified for models 
like the spinning Lorentz gas, which is deterministic, chaotic and interacting, 
indicates further the more universal validity of 
this class of {\it active} fluctuation symmetries.\\

\noindent {\bf Acknowledgment} Financial support in the form of a Belgian InterUniversity Grant from Belspo
is gratefully acknowledged.

\bibliographystyle{plain}

\begin{thebibliography}{10}


\bibitem{em}
D.J.~Evans, E.~G.~D.~Cohen and G.P.~Morris, Probability of second law violations in
steady flows.  Phys. Rev. Lett. {\bf 71}, 2401--2404
(1993).


\bibitem{gc} G.~Gallavotti and E.~G.~D.~Cohen, Phys. Rev. Lett.
\textbf{74}, 2694 (1995); J. Sat. Phys. \textbf{80}, 931 (1995).



\bibitem{maes}
C.~Maes, The fluctuation theorem as a Gibbs property. J. Stat.
Phys. {\bf 95}, 367--392 (1999).

\bibitem{LS}
J.L~Lebowitz and H.~Spohn, The Gallavotti--Cohen fluctuation theorem for stochastic
dynamics. J. Stat. Phys. {\bf 95}, 333 (1999).

\bibitem{crooks}
G.~E.~Crooks, Entropy production fluctuation theorem and the nonequilibrium work relation for free energy
differences. Phys. Rev. E {\bf 60}, 2721--2726 (1999).


\bibitem{time}
C.~Maes and K.~Neto\v{c}n\'y, Time-reversal and Entropy. J. Stat. Phys. {\bf 110}, 269--310 (2003).

\bibitem{poincare}
C.~Maes, On the origin and the use of fluctuation relations for the entropy. S\'eminaire Poincar\'e {\bf 2},  29--62 
Eds. J. Dalibard, B. Duplantier and V. Rivasseau, Birk\"{a}user (Basel), 2003.

\bibitem{leb}
P.G.~Bergman and J.L.~Lebowitz, New Approach to Nonequilibrium Process. Phys. Rev. {\bf 99}, 578-587, 1955.

\bibitem{kls}
S.~Katz, J.L.~Lebowitz, and H.~Spohn, Stationary nonequilibrium states for stochastic lattice gas models of
ionic superconductors. J. Stat. Phys. {\bf 34}, 497--538 (1984).


\bibitem{har}
T.~Harada and S.-Y.~Sasa: Equality connecting energy
dissipation with violation of fluctuation-response relation, {\it
Phys. Rev. Lett.} {\bf 95}, 130602 (2005).

\bibitem{der}
B.~Derrida, Non-equilibrium steady states: fluctuations and large deviations of the density and of the
current. J. Stat. Mech. P07023 (2007).


\bibitem{hal}
H.~Tasaki, Two theorems that relate discrete stochastic processes to microscopic mechanics.
arXiv:0706.1032v1 [cond-mat.stat-mech]. 

\bibitem{jarz}
C.~Jarzynski, Phys. Rev. Lett. \textbf{78}, 2690 (1997);
Phys. Rev. E \textbf{56}, 5018 (1997).


\bibitem{maarten}
C.~Maes and M.H.~van Wieren, Time-symmetric fluctuations in nonequilibrium systems. Phys. Rev. Lett. \textbf{96}, 240601 (2006). 

\bibitem{chan}
R.~Jack, J.P.~Garrahan, D.~Chandler: Space-time thermodynamics and subsystem observables in kinetically constrained models of glassy materials,
\emph{J. Chem. Phys.} \textbf{125}, 184509 (2006).

\bibitem{vW}
J.~P.~Garrahan, R.L.~Jack, V.~Lecomte, E.~Pitard, K.~van
Duijvendijk, and F.~van Wijland: First-order dynamical phase
transition in models of glasses: an approach based on ensembles of
histories,
\emph{J.~Phys. A: Math. Gen.} {\bf 42}, 075007 (2009).

\bibitem{woj}
W.~De Roeck and C.~Maes, Symmetries of the ratchet current. Phys. Rev. E {\bf 76}, 051117 (2007).

\bibitem{fdr}
M.~Baiesi, E.~Boksenbojm, C.~Maes and B.~Wynants,  Nonequilibrium linear response for Markov dynamics, I: jump processes and overdamped diffusions, J. Stat .Phys. {\bf 137}, 1094-1116 (2009).

\bibitem{kubo}
R.~Kubo, The fluctuation-dissipation theorem. Rep. Prog. Phys. {\bf 29}, 255-–284 (1966).

\bibitem{mat}
M.~Colangeli, C.~Maes and B.~Wynants, A meaningful expansion around detailed balance. J. Phys. A: Math. Theor. {\bf 44}, 095001 (2011).


\bibitem{seki}
K.~Sekimoto, {\it Stochastic Energetics}. Lecture Notes in Physics {\bf 799}, Springer (2010).

\bibitem{rito}
F.~Ritort, C.~Bustamente and I.~Tinoco,Jr., A two-state kinetic model for
the unfolding of single molecules by mechanical force. Proc. Nat. Ac. Scienc. {\bf 99}, 13544-–13538
(2002).

\bibitem{har07} R.~J.~Harris and G.~M.~Sch\"{u}tz, Fluctuation theorems for stochastic dynamics, J.
Stat. Mech: Theory and Experiment, P07020 (2007).

\bibitem{kurch}
J.~Kurchan, Fluctuation theorem for stochastic dynamics. J. Phys. A: Math. Gen. {\bf 31}, 3719-–3729 (1998).

\bibitem{jmp}
C.~Maes, F.~Redig and A.~Van Moffaert, On the definition of entropy production, via examples. J. Math. Phys. {\bf 41}, 1528-1554 (2000). 


\bibitem{ruelle}
D.~Ruelle, Smooth Dynamics and New Theoretical Ideas in Nonequilibrium Statistical Mechanics.
J. Stat. Phys. {\bf 95}, 393-–468 (1999).

\bibitem{verb}
C.~Maes and E.~Verbitskyi, 
Large deviations and a fluctuation symmetry for chaotic homeomorphisms.
Comm. Math. Phys. {\bf 233}, 137--151 (2003). 

\bibitem{an}
D.~Andrieux and P.~Gaspard, The fluctuation theorem for currents in semi-Markov processes. J. Stat. Mech.:Theory and Experiment,  P11007 (2008). 


\bibitem{bon}
F.~Bonetto, G.~Gallavotti, and P.~Garrido, Chaotic principle: An experimental test.
Physica D {\bf 105}, 226 (1997).


\bibitem{semim}
C.~Maes, K.~Neto\v{c}n\'y and B.~Wynants, Dynamical fluctuations for semi-Markov processes. J. Phys. A: Math. Theor. {\bf 42}, 365002 (2009). 


\bibitem{gal}
G.~Gallavotti, Extension of Onsager’s Reciprocity to Large Fields and the Chaotic Hypothesis. Phys. Rev. Lett. {\bf 77}, 4334--4337 (1996).


\bibitem{koma}
T.~S.~Komatsu and N.~Nakagawa, An expression for stationary distribution in nonequilibrium
steady states. Phys. Rev. Lett. {\bf 100}, 030601 (2008).


\bibitem{mcl}
C.~Maes and  K.~Neto\v{c}n\'y, Rigorous meaning of McLennan ensembles. J. Math. Phys. {\bf 51}, 015219 (2010).

\bibitem{beyond}
C.~Maes, K.~Neto\v{c}n\'y, and B.~Wynants,  On and beyond entropy production; the case of Markov jump processes. Markov Proc. Rel. Fields. {\bf 14}, 445--464 (2008).

\bibitem{prague}
C.~Maes,  K.~Neto\v{c}n\'y and B.~Shergelashvili, A selection of nonequilibrium issues. In: Methods of Contemporary Mathematical Statistical Physics, Ed. Roman Koteck\'y, Lecture Notes in Mathematics 1970, pp. 247-306, Springer, 2009. 

\bibitem{rose}
R.~J.~Harris, A.~R\'akos and G.M.Sch\"utz, 
Breakdown of Gallavotti-Cohen symmetry for stochastic dynamics. Europhys. Lett. {\bf 75}, 227 (2006).

\bibitem{Lor05}  H.~A.~Lorentz, Proc. Amst. Acad. {\bf 438} (1905).

\bibitem{Gas98} P.~Gaspard, {\it Chaos, Scattering, and Statistical Mechanics} (Cambridge University Press,
Cambridge, 1998).

\bibitem{Sza00} {\it Hard Ball Systems and the Lorentz Gas}, edited by D. Sz\'{a}sz (Springer Verlag, Berlin, 2000).

\bibitem{Lar03} H.~Larralde, F.~Leyvraz and C.~Mejia-Monasterio, Transport properties of a modified Lorentz gas.
J. Stat. Phys. {\bf 113}, 197--231 (2003).

\bibitem{Sal09} A.~Salazar, F.~Leyvraz and H.~Larralde, Fluctuation theorem for currents in the Spinning Lorentz Gas.
Physica A \textbf{388}, 4679--4694 (2009). 

\bibitem{Reif65} F.~Reif, {\it Fundamentals of Statistical and Thermal Physics} 
(International Student Edition), McGraw-Hill, Tokyo, 1965; pp. 272.

\end{thebibliography}

\end{document}